\documentclass[10pt,conference]{IEEEtran}

\usepackage[ruled,vlined,linesnumbered,noend]{algorithm2e}
\usepackage{algorithmic}
\usepackage{tabularray}
\usepackage{threeparttable}
\usepackage{graphicx}
\usepackage[percent]{overpic}
\usepackage{booktabs}
\usepackage{array}
\usepackage{amsmath}
\usepackage{amsthm}
\usepackage[compatibility=false]{caption}
\usepackage{subcaption}
\usepackage[most]{tcolorbox}
\usepackage{enumitem}
\usepackage{hyperref}
\usepackage{listings}
\usepackage{xcolor}
\usepackage{xspace}
\usepackage{rotating}
\usepackage{dblfloatfix}

\AtBeginDocument{%
  \providecommand\BibTeX{{%
    \normalfont B\kern-0.5em{\scshape i\kern-0.25em b}\kern-0.8em\TeX}}}

\newcommand*{\tool}{{\textsc{APG}}\xspace}

\theoremstyle{definition}

\newcounter{definitioncounter}

\newcounter{examplecounter}

\newtcolorbox{formulateBox}[1][]{colback=white,
   boxrule=0.2mm,
   arc=0mm,
   width=\columnwidth,
   left=0.2mm,
   right=0.2mm,
   top=0.2mm,
   bottom=0.2mm,
   #1}

\newtcolorbox{answerBox}[1][]{colback=gray!10,
   colframe=gray!10,
   boxrule=0.2mm,
   arc=0.5mm,
   width=\columnwidth,
   left=0.2mm,
   right=0.2mm,
   top=0.2mm,
   bottom=0.2mm,
   #1}

\hypersetup{
  pdftitle={APG: Agentic Static-Analysis Infrastructure on LLVM-IR Graph Databases},
  pdfauthor={Zexin Zhong, Yanjun Zhang, Yulei Sui, Xiao Cheng}
}

\begin{document}

\title{\tool{}: Agentic Static-Analysis Infrastructure on LLVM-IR Graph Databases}

\author{
\IEEEauthorblockN{Zexin Zhong}
\IEEEauthorblockA{
University of Technology Sydney\\
Sydney, Australia\\
zexin.zhong@student.uts.edu.au
}
\and
\IEEEauthorblockN{Yanjun Zhang}
\IEEEauthorblockA{
Griffith University\\
Queensland, Australia\\
yanjun.zhang@griffith.edu.au
}
\and
\IEEEauthorblockN{Yulei Sui}
\IEEEauthorblockA{
University of New South Wales\\
Sydney, Australia\\
y.sui@unsw.edu.au
}
\and
\IEEEauthorblockN{Xiao Cheng}
\IEEEauthorblockA{
Macquarie University\\
Sydney, Australia\\
xiao.cheng@mq.edu.au
}
}

\maketitle

\begin{abstract}
LLM-based coding agents increasingly rely on external program
knowledge, yet access to static-analysis graphs alone does not
ensure appropriate operation selection or evidence interpretation.
We present \tool, an infrastructure for
semantics-grounded static program analysis.
Built on SVF and LLVM, it decomposes compiler-derived program
analyses into atomic capabilities that LLM agents can select,
compose, and orchestrate for goal-directed, cross-graph
investigation, with query safeguards and evidence traces
preserving inspectable results and provenance.
\tool provides 38 atomic capabilities.
On analysis requests arising during benchmark repair runs,
35 achieve first-attempt query-generation success rates above
50\%, including 13 at 100\%.
Evaluation on 812 C/C++ defects shows higher repair success
than the evaluated baselines.
Compared with Passive Graph Push, \tool improves
first-iteration repair success by 17.6 and 18.0 percentage points
on the two benchmark collections and reduces average LLM
input-token consumption by 90.8\%.
Our tool is available at
\url{https://github.com/JasonZhongZexin/APG-ICSE-Demo} and its demonstration video at
\url{https://youtu.be/vkuExZUjGnw}.
\end{abstract}

\section{Introduction}
\label{sec:introduction}

Interprocedural static analysis on LLVM IR,
as realized by frameworks such as
SVF~\cite{sui2016svf}, computes precise
structural facts about system-level C/C++
programs, including pointer aliasing,
value-flow dependencies, and control-flow
reachability across compilation units.
Such structural evidence can inform fault
localization~\cite{wong2016survey}
and program repair~\cite{fan2024apr},
yet providing agents with usable access
to these analyses remains a challenge.
RLCE retrieves repository context for
repair~\cite{chen2024rlce}; STALL+ integrates
static analysis into code
completion~\cite{liu2024stall}; and RepoGraph
provides graph-based repository
navigation~\cite{ouyang2025repograph}.
These systems focus on repository context,
code completion, and navigation.
Exposing LLVM-IR interprocedural analyses
as reusable, agent-invocable capabilities
poses a further infrastructure challenge.

These facts should instead be supplied
to agents directly, letting the model
focus on higher-level reasoning
such as defect interpretation,
root-cause localization, and fix
synthesis. A graph database is a
natural vehicle: many program analyses
reduce to graph
reachability~\cite{reps1997program},
and a declarative language such as
Cypher lets an agent express \emph{what}
structural information it needs
without encoding \emph{how} to traverse
the analyzer's internal data structures,
a separation far more amenable to LLM
interaction than imperative analysis APIs.

However, \emph{merely connecting an LLM
to a static-analysis database does not
make it capable of using static analysis
correctly}. Program-analysis graphs encode
specialized relationships whose meanings
depend on their underlying analysis
semantics. An agent must determine which
analysis operation is appropriate, how to
express its intended goal, and what
conclusions the returned evidence supports.
A schema-valid query may still investigate
the wrong relationship, while an empty
result does not necessarily establish the
absence of a program property.

This gap motivates two design principles.
First, program-analysis infrastructure
should be \emph{agent-facing}: designed for
LLM agents as primary consumers, exposing
semantically meaningful analysis
capabilities rather than only low-level
graph operations. Second, it should be
\emph{inspectable}: agent-generated
operations should be checked against
supported constraints, and the evidence
underlying analysis conclusions should
remain traceable.

We present \tool, an infrastructure for
semantics-grounded static program analysis.
Built on SVF and LLVM, \tool decomposes
compiler-derived program analyses into
atomic, agent-invocable capabilities
through a semantics-aware interface.
LLM agents select and orchestrate these
capabilities for goal-directed, iterative
investigation of program structures, while
query safeguards and provenance-aware
evidence organization support constrained
and inspectable program investigation.

The main tool contributions are:
\begin{itemize}
    \item \textbf{LLVM-IR graph database
    infrastructure.} \tool materializes
    five SVF-derived interprocedural analysis
    graphs (PAG, ICFG, Call Graph, SVFBasicBlock
    Graph, and CHG) into a queryable graph
    database, making compiler-precise program
    relationships accessible via declarative
    Cypher queries.
    \item \textbf{Agent-facing analysis interaction
    and orchestration.} \tool decomposes
    static analyses into atomic capabilities
    that LLM agents select and orchestrate
    through a semantics-aware interface,
    enabling goal-directed, iterative
    investigation of program structures.

    \item \textbf{Inspectable, evidence-grounded
    investigation.} \tool integrates
    query safeguards and provenance-aware
    evidence organization to make analysis
    operations and their supporting structural
    evidence inspectable.
\end{itemize}

\section{Tool Design and Implementation}
\label{sec:codegraphhunter}

\begin{figure*}[!t]
    \centering
    \input{Figures/Fig1-APG-links.tex}
    \vspace{-5mm}
    \caption{Overview of \tool.}
    \vspace{-3mm}
    \label{fig:architecture}
\end{figure*}

Fig.~\ref{fig:architecture} illustrates
the architecture of \tool. Building on
SVF~\cite{sui2016svf}, the framework
materializes five compiler-derived
program-analysis graphs (PAG, ICFG,
Call Graph, SVFBasicBlock Graph, and CHG)
in TuGraph~\cite{alibaba2024tugraph}
for on-demand structural queries.

Given a defect description or analysis
question, the agent formulates analysis
goals and selects semantic capabilities
through a semantics-aware interface.
An iterative controller coordinates
query assessment, execution, and
follow-up investigation. Retrieved
results are consolidated into local
and global evidence summaries, while
analysis traces preserve the operations,
queries, and execution feedback for
inspection. The accumulated evidence
supports fault diagnosis and downstream
automated repair.

The architecture realizes two principles:
\emph{agent-facing interaction}, which
exposes static-analysis operations as
agent-invocable semantic capabilities,
and \emph{inspectable investigation},
which constrains operations and retains
traceable evidence. The following
subsections describe the three mechanisms.

\subsection{Semantics-Grounded Analysis Capabilities}
\label{sec:semantic_interface}

Rather than exposing program graphs solely
as schemas, the semantics-aware interface
interposes a semantic layer between an
agent's analysis intention and its
executable graph operation. It comprises
a Graph Semantics Catalog, an Atomic
Capability Catalog, and reusable
graph-query operators.

The design separates three levels:
\emph{analysis intent} (the structural
information being sought),
\emph{semantic capability} (the
corresponding program-analysis operation),
and \emph{query realization} (expressing
that operation over the graph).
The Graph Semantics Catalog describes
graph entities, relationships, and their
analysis meanings. Atomic capabilities
provide reusable abstractions for
selecting meaningful operations.

This separation is necessary because
schema validity alone does not establish
analysis correctness. For example,
determining whether function $f$ can
transitively call $g$ requires a
reachability traversal, not a
direct-edge lookup. The semantic layer
thus guides both operation selection
and evidence interpretation.

Each capability is exposed as an invocable interface.
Given the selected capability and the current analysis intent,
the LLM generates Cypher using the associated query templates
and reusable Cypher operators.
Query Guard checks syntax, safety constraints, and compatibility
with the query features supported by the deployed TuGraph engine.
Schema Guard checks whether the node and edge properties
referenced by the query are defined in the graph schema.
Goal/query alignment is assessed separately during investigation.

\begin{table*}[t]
\centering
\caption{Capability coverage and capability-level Query S@1.
Interval columns report numbers of capabilities.}
\vspace{-2mm}
\label{tab:capabilities}
\small
\setlength{\tabcolsep}{5pt}
\begin{tabular}{lrrrrp{0.43\textwidth}}
\hline
& & \multicolumn{3}{c}{\textbf{Query S@1}} & \\
\textbf{Category} &
\textbf{Count} &
\textbf{100\%} &
\textbf{$(50,100)$\%} &
\textbf{$\leq$50\%} &
\textbf{Representative capabilities} \\
\hline
CallGraph & 8 & 3 & 5 & 0 &
Call chains; reachability; self-recursion; mutual recursion \\

ICFG & 8 & 2 & 5 & 1 &
Branch context; guard candidates; interprocedural paths;
loop detection \\

PAG & 10 & 4 & 5 & 1 &
Argument flow; load/store uses; parameter passing;
approximate alias evidence \\

CHG/Type & 5 & 0 & 4 & 1 &
Class hierarchy; virtual-dispatch candidates;
override relations; type/field information \\

Cross-graph & 7 & 4 & 3 & 0 &
Entity mapping; evidence aggregation; check-before-use;
sanitizer-to-sink investigation \\
\hline
\textbf{Total} & \textbf{38} & \textbf{13} &
\textbf{22} & \textbf{3} &
Excludes five source/symbol capabilities. \\
\hline
\end{tabular}
\vspace{-2mm}
\end{table*}

\begin{table}[t]
\centering
\caption{Query reliability and repair success (\%).}
\vspace{-2mm}
\label{tab:query-ablation}
\small
\begin{tabular}{lrr}
\hline
\textbf{Configuration} &
\textbf{IQR $\downarrow$} &
\textbf{Repair S@1 $\uparrow$} \\
\hline
Without Graph Semantics Catalog & 68.4 & 4.8 \\
Without Guards/Replanning & 14.2 & 21.5 \\
Full Pipeline             & 0.8  & 32.8 \\
\hline
\end{tabular}
\vspace{-2mm}
\end{table}

\begin{table}[t]
\centering
\caption{Repair success within one and five sequential
iterations (\%).}
\vspace{-2mm}
\label{tab:repair-results}
\small
\setlength{\tabcolsep}{3pt}
\begin{tabular}{lrrrr}
\hline
& \multicolumn{2}{c}{\textbf{Multi-SWE}}
& \multicolumn{2}{c}{\textbf{D4C/ManyBugs}} \\
\textbf{Method}
& \textbf{S@1} & \textbf{S@5}
& \textbf{S@1} & \textbf{S@5} \\
\hline
Text-RAG        & 3.9  & 8.2  & 2.4  & 5.8 \\
AST-Agent       & 13.8 & 20.4 & 8.9  & 14.5 \\
Passive Push    & 15.2 & 24.1 & 11.2 & 18.4 \\
\tool & 32.8 & 45.4 & 29.2 & 41.6 \\
\hline
\end{tabular}
\vspace{-2mm}
\end{table}

\subsection{Goal-Directed Structural Investigation}
\label{sec:active_exploration}

\tool organizes exploration as a
sequence of explicit analysis goals
and evidence-acquisition steps.
An \texttt{AttemptController} coordinates
this process under bounded execution
budgets, so the agent progressively
investigates program relationships
rather than retrieving a fixed
collection of graph context.

At each step, the agent identifies an information need,
selects a capability, and generates a candidate Cypher query.
Before execution, the query undergoes goal/query alignment
assessment, Query Guard, and Schema Guard.
Alignment assessment examines whether the query expresses
the intended analysis goal, beyond syntax and schema validity.
Only queries passing all applicable checks proceed to execution;
otherwise, the agent revises the query or replans the investigation.

Returned results are interpreted against
the current goal. A zero-hit result
establishes only that no matching evidence
was found within the queried scope. It
may reflect an absent relationship,
incomplete graph information, or an
undetected query-goal mismatch.

Based on accumulated evidence, the agent
determines whether further investigation
is needed. For example, diagnosing a
cross-file memory defect may require
combining call-graph, pointer, and
control-flow evidence from different
graphs. Each retrieval step is thus tied
to an explicit analysis objective, unlike
passive graph-context injection.

\subsection{Inspectable Evidence-Grounded Reasoning}
\label{sec:whitebox_reasoning}

\tool separates retrieved program
evidence from the conclusions the LLM
derives from it. Individual query
results are processed into local
summaries of relevant entities and
structural relationships. A global
summary consolidates evidence across
investigation steps, providing
structured context for subsequent
reasoning without requiring the full
collection of raw query outputs.

Crucially, summarization does not
replace the underlying records.
\tool maintains analysis traces
and provenance linking operations,
executed queries, graph results,
and execution feedback. Users can
therefore inspect the evidence
available to the agent and examine
potential failures in the investigation
process.
The same investigation mechanism
supports multiple downstream tasks:
structural program analysis, fault
localization and root-cause diagnosis,
and automated repair (which extends
the process with patch generation
and test feedback).
The resulting \emph{white-box} property
concerns the inspectability of external
analysis operations and their supporting
evidence, not visibility into the LLM's
internal computations or formal
correctness guarantees.

\section{Evaluation}
\label{sec:evaluation}

We evaluate \tool on
\textbf{RQ1: capability coverage and query generation},
\textbf{RQ2: query reliability}, and
\textbf{RQ3: repair effectiveness and retrieval efficiency}.

\subsection{Experimental Setup}

We use 812 C/C++ defects: 277 from
Multi-SWE-bench~\cite{zan2025multiswebenchmultilingualbenchmarkissue}
and 535 from Defects4C~\cite{defects4c2025}
and ManyBugs~\cite{legoues2015manybugs}.
Baselines are Text-RAG (source retrieval),
AST-Agent (syntax-level context), and
Passive Graph Push (preassembled graph context).
All configurations use GPT-4o with temperature~0.0
and a clean checkout for each defect, allowing
up to five sequential repair iterations.
Each iteration includes context acquisition,
patch generation, and compilation/test validation.
Repair Success@$k$ measures success within $k$
sequential iterations, rather than independently
sampled Pass@$k$.

\subsection{RQ1: Capability Coverage and Query Generation}

\tool provides 38 atomic analysis capabilities
across five graph categories
(Table~\ref{tab:capabilities}).
For each capability, \emph{Query S@1} measures
the proportion of analysis requests whose initial
Cypher query correctly expresses the intended goal.
Corrections and retries do not count as
first-attempt successes.

Of the 38 capabilities, 35 achieve Query S@1
above 50\%, including 13 at 100\%.
These rates characterize the benchmark workload,
rather than guaranteeing the same outcomes
for unseen requests.
Capabilities compose through shared entities
and source locations.
For example, parameter propagation identifies
a use location for branch-context retrieval,
while call-chain exploration supplies its
calling context.

\subsection{RQ2: Query Reliability}

We ablate the Graph Semantics Catalog and the
combined guards/replanning mechanisms.
Invalid Query Rate (IQR) measures the proportion
of query observations containing Cypher syntax
or graph-schema violations.

As shown in Table~\ref{tab:query-ablation},
the full pipeline reduces IQR to 0.8\% and
achieves the highest repair S@1.
The results support the role of the Graph
Semantics Catalog and query safeguards in
maintaining query validity.

\subsection{RQ3: Repair Effectiveness and Retrieval Efficiency}

\tool achieves the highest S@1 and S@5 on both
benchmarks (Table~\ref{tab:repair-results}).
Compared with Passive Graph Push, S@1 improves
by 17.6 and 18.0 percentage points on
Multi-SWE-bench and Defects4C/ManyBugs.
Under the same accounting scope, average LLM
input tokens per defect drop from 45.8K
(Passive Push) to 4.2K (\tool), a 90.8\%
reduction. This is consistent with acquiring
structural evidence on demand rather than
injecting preassembled graph context.

\section{Related Work}
\label{sec:related_work}

\noindent\textbf{Repository-Level Agent Interaction.}
SWE-agent enables LLM agents to navigate,
edit, and test code repositories~
\cite{yang2024sweagent}. RepoCoder and
RLCE improve repository-level context
retrieval for code generation and repair~
\cite{zhang2023repocoder,chen2024rlce},
while RepoGraph provides graph-based
repository navigation~
\cite{ouyang2025repograph}.
These approaches facilitate repository
exploration. \tool addresses
a complementary challenge: enabling
agents to invoke program-analysis
operations with explicit semantics,
rather than relying solely on
retrieved repository context.

\noindent\textbf{Structural Knowledge Integration.}
STALL+ integrates static-analysis
information into repository-level code
completion~\cite{liu2024stall}, while
CGM incorporates code graphs into an
LLM's attention mechanism~\cite{tao2025cgm}.
More closely related, CT-Repair
enables agents to query static Code
Property Graphs and dynamic Temporal
Execution Graphs to develop
evidence-grounded repair strategies~
\cite{huang2026ctrepair}.
Unlike model-integrated graph
representations or repair-specific
evidence workflows, \tool
emphasizes reusable semantic analysis
capabilities exposed through an
external, agent-facing interface.

\noindent\textbf{Agentic Program-Analysis Querying.}
Declarative query languages support
expressive program analysis~
\cite{de2008ql}. QLCoder addresses
LLM interaction with static analysis
by synthesizing CodeQL vulnerability
queries using syntax guidance and
execution feedback~\cite{wang2026qlcoder}.
\tool targets a different abstraction:
mapping analysis intentions to semantic
capabilities and executable graph queries,
then organizing multi-round investigation
around inspectable structural evidence
and provenance.

\section{Conclusion}

We presented \tool, an infrastructure that materializes
five SVF-derived interprocedural analysis graphs into a
queryable graph database and decomposes the resulting
static analyses into 38 atomic, agent-invocable capabilities.
LLM agents select and orchestrate these capabilities through
a semantics-aware interface for goal-directed, cross-graph
program investigation, while query safeguards and
provenance-aware evidence organization keep the investigation
process constrained and inspectable.

\bibliographystyle{IEEEtran}
\bibliography{reference}

\end{document}